\documentclass[twoside,11pt]{article}

\usepackage[accepted]{melba}

\usepackage{amsmath,amsfonts}

\usepackage{booktabs}
\usepackage{multirow}
\usepackage{longtable}
\usepackage{makecell}
\usepackage{float}
\usepackage{caption}
\usepackage{subcaption}
\usepackage{xcolor}
 \usepackage{array,graphicx}
\usepackage{bm}

\melbaheading{2022:031}{https://www.melba-journal.org/papers/2022:031.html}{2022}{1-27}{03/2022}{01/2023}{Lemay, Gros, Naga Karthik and Cohen-Adad}{}{}

\ShortHeadings{Label fusion and training methods for inter-rater uncertainty}{Lemay, Gros, Naga Karthik, and Cohen-Adad}
\firstpageno{1}

\title{Label fusion and training methods for reliable representation of inter-rater uncertainty}

\author{\name Andreanne Lemay \email andreanne.lemay@polymtl.ca \\  
	\addr NeuroPoly Lab, Institute of Biomedical Engineering, Polytechnique Montreal, Montreal, QC, Canada \\
	\addr Mila, Quebec AI Institute, Montreal, QC, Canada \\
	\AND
	\name Charley Gros \email charley.gros@gmail.com \\
	\addr NeuroPoly Lab, Institute of Biomedical Engineering, Polytechnique Montreal, Montreal, QC, Canada \\
	\addr Mila, Quebec AI Institute, Montreal, QC, Canada \\
	\AND
 	\name Enamundram Naga Karthik \email naga-karthik.enamundram@polymtl.ca \\
	\addr NeuroPoly Lab, Institute of Biomedical Engineering, Polytechnique Montreal, Montreal, QC, Canada \\
	\addr Mila, Quebec AI Institute, Montreal, QC, Canada \\
    \AND
	\name Julien Cohen-Adad \email jcohen@polymtl.ca \\
	\addr NeuroPoly Lab, Institute of Biomedical Engineering, Polytechnique Montreal, Montreal, QC, Canada \\
	\addr Mila, Quebec AI Institute, Montreal, QC, Canada \\
	\addr Functional Neuroimaging Unit, CRIUGM, Université de Montréal, Montreal, QC, Canada\\
}

\begin{document}

\maketitle

\vspace{-3em}
\begin{abstract}
    Medical tasks are prone to inter-rater variability due to multiple factors such as image quality, professional experience and training, or guideline clarity. Training deep learning networks with annotations from multiple raters is a common practice that mitigates the model’s bias towards a single expert. Reliable models generating calibrated outputs and reflecting the inter-rater disagreement are key to the integration of artificial intelligence in clinical practice. Various methods exist to take into account different expert labels. We focus on comparing three label fusion methods: STAPLE, average of the rater’s segmentation, and random sampling of each rater’s segmentation during training. Each label fusion method is studied using both the conventional training framework and the recently published SoftSeg framework that limits information loss by treating the segmentation task as a regression. Our results, across 10 data splittings on two public datasets (spinal cord gray matter challenge, and multiple sclerosis brain lesion segmentation), indicate that SoftSeg models, regardless of the ground truth fusion method, had better calibration and preservation of the inter-rater rater variability compared with their conventional counterparts without impacting the segmentation performance. Conventional models, i.e., trained with a Dice loss, with binary inputs, and sigmoid/softmax final activate, were overconfident and underestimated the uncertainty associated with inter-rater variability. Conversely, fusing labels by averaging with the SoftSeg framework led to underconfident outputs and overestimation of the rater disagreement. In terms of segmentation performance, the best label fusion method was different for the two datasets studied, indicating this parameter might be task-dependent. However, SoftSeg had segmentation performance systematically superior or equal to the conventionally trained models and had the best calibration and preservation of the inter-rater variability. SoftSeg has a low computational cost and performed similarly in terms of uncertainty to ensembles which require multiple models and forward passes. Our code is available at~\url{https://ivadomed.org}.
\end{abstract}


\section{Introduction}
Manual annotation of medical images is challenged by ill-defined boundaries between anatomical regions, and hence prone to inter-expert variability. Inter-expert disagreement is widely acknowledged as a key limitation in medical image analysis \citep{schaekermann2019understanding} as it hinders the definition of ground truth (GT) annotation \citep{mirikharaji2021d, shwartzman2019worrisome, yu2020difficulty}. For instance, the multiple sclerosis (MS) brain dataset annotated by 7 experts reported an inter-expert agreement ranging between experts from 0.66 to 0.76 of median Dice score with the consensus \citep{commowick2021multiple}. This variability can arise from many factors, including image quality, expert experience, or guidelines clarity \citep{mirikharaji2021d, schaekermann2019understanding}. To mitigate this issue along with speeding annotating time and enhancing reproducibility, a large number of automatic annotation algorithms have been proposed \citep{gabr2020brain, gros2019automatic, isensee2017brain, lemay2021automatic}. However, the annotations provided by these automatic algorithms are likely to reflect the characteristics of the data they are trained on, including the biases they carry such as different expert experience or style \citep{vincent2021impact}. Therefore, it is common practice to provide, for each image, annotations from multiple experts \citep{commowick2018objective, mirikharaji2021d, prados2017spinal, schaekermann2019understanding}. It remains, however, unclear how to properly use these multiple experts’ annotations, i.e., to combine them to generate a GT, to preserve the inter-rater variability information while limiting the expert bias encoded in the model \citep{mirikharaji2021d}.

\subsection{Study outline}
This study compares different methods to aggregate multiple experts' annotations as GT in algorithm training. A common method to use multiple experts’ annotations is to fuse them to create a single mask per image. The fusion method can lead to masks with either categorical values (e.g., zeros or ones for a one-class segmentation task) or soft values (e.g., between 0 and 1), hereafter called “hard fusion” and “soft fusion”, respectively. Hard fusion methods include “Simultaneous truth and performance level estimation” (STAPLE) \citep{warfield2004simultaneous}, majority voting, intersection, or union, and were widely used in the automatic segmentation literature. On the other hand, soft fusion methods, e.g., averaging the experts’ annotations, received a modest interest, probably because most segmentation algorithms assume GTs with categorical values. A training pipeline, called SoftSeg, has been recently proposed to favor the propagation of soft labels (i.e., non-categorical values) \citep{gros2021softseg}. The comparison between soft and hard fusion methods questions the tradeoff between the precision and the generalization of a “gold standard” as a precise ground-truth (i.e., hard / binary) may not be reflective of the underlying inter-expert uncertainty. Alternatively, one can choose not to fuse the experts’ annotations and, instead, to use them independently when training a segmentation method. This approach is hereafter called random sampling method and aims to preserve the raw multi-expert labeling while confronting the algorithm to contradictory annotations \citep{jensen2019improving, jungo2018effect, mirikharaji2021d}.

\subsection{Related works}
Some studies compared methods to generate GT labels when experts disagree. Jensen et al. demonstrated that hard fusion, i.e., majority voting, led to over-confident models on skin disease predictions (i.e., uncalibrated model) \citep{jensen2019improving}. They showed that a “no fusion” approach, i.e., label random sampling, could mitigate this miscalibration in the model’s prediction. Jungo et al. compared hard fusion (STAPLE, majority voting, intersection, and union) methods with the random sampling approach in terms of segmentation performance and uncertainty estimation \citep{jungo2018effect}. The random sampling method yielded uncertainty that was able to reflect the underlying expert disagreement on synthetic data and on subjects with a Dice score superior to the median of a brain tumor dataset, but no positive impact was noticed for subjects with a low segmentation performance. Conversely, the hard fusion methods led to an under-estimation of uncertainty, suggesting that inter-expert variability needs to be explicitly taken into consideration when training models in order to reliably estimate uncertainty. To the best of our knowledge, there is currently no study that compares soft fusion methods with hard fusion and random sampling approaches.

Increasing attention has been given to uncertainty estimation. Notably, Quantification of Uncertainties in Biomedical Image Quantification Challenge (QUBIQ) aimed at developing uncertainty estimation methods that were evaluated on MRI and CT scan datasets with multiple raters per annotation. The QUBIQ2020 winning team proposed training one model per rater and aggregating the outputs to evaluate the segmentation uncertainty \citep{ma2020estimating}. \cite{silva2021using} proposed to use soft labels computed by averaging all labels and had well-calibrated outputs, competitive with other approaches from the QUBIQ2021 challenge.

\subsection{Our contribution}
In this study, we compare the impact of hard fusion, soft fusion, and label random sampling methods using SoftSeg or a conventional training framework. The inter-rater variability is lost in hard fusion methods \citep{yu2020difficulty} and the conventional framework, which inputs binarized GT and trains with categorical losses, limiting the learning of expert disagreement. Hence, we hypothesize that soft or random sampling methods and the SoftSeg framework will better reflect the inter-rater variability, will generate more calibrated predictions, and will yield improved segmentation performances than hard fusion and conventional training methods. The label generated by these methods is used to feed a U-Net \citep{ronneberger2015u}, widely considered as the state-of-the-art in automatic image segmentation. The training is performed using both a conventional pipeline and the recently proposed alternative, SoftSeg. Each method, six in total (two training pipelines, each using the three methods to generate the GT, see Table \ref{tab:candidate}), are compared on two MRI open-source datasets, the spinal cord gray matter (SCGM) challenge \citep{prados2017spinal} and multiple sclerosis (MS) brain lesion challenge \citep{commowick2018objective}, in terms of (i) preservation of the inter-rater variability, (ii) model calibration and, (iii) segmentation performance.

\section{Method and Material}
\subsection{Method}
\subsubsection{Label fusion}
Three methods to exploit multiple rater labels were studied: STAPLE \citep{warfield2004simultaneous}, average across GTs, and random sampling of one annotation during training without fusion \citep{jungo2018effect}. STAPLE is an expectation-maximization algorithm widely used for label fusion in medical imaging \citep{akkus2017deep, commowick2018objective, maier2017isles}. This method produces binary GTs. The second label fusion strategy studied, averaging across all annotations, aims to preserve all the inter-rater variability information by outputting soft (i.e., values between 0 and 1) GTs. However, conventional segmentation pipelines usually binarize the GTs which leads to a majority voting when averaging segmentations across raters. To fully exploit this label fusion method, a soft segmentation framework such as SoftSeg \citep{gros2021softseg} is required. The third method does not merge the labels. During each training epoch, one rater segmentation is randomly chosen as GT, eventually exposing the model to all the rater’s annotations. This means that all training cases from all raters are shown to the model during the training. Therefore, the random sampling method uses binary segmentations. 

\subsubsection{Training framework}
In this work, we compare each label fusion method when trained with both SoftSeg and a conventional segmentation training framework. SoftSeg has three differences when compared with the conventional approach: no binarization during the preprocessing and data augmentation, soft final activation function, and training using a regression loss function \citep{gros2021softseg}. The final activation and the regression loss function are normalized ReLU and Adaptive Wing loss \citep{wang2019adaptive} respectively as defined in \cite{gros2021softseg}. The final activation was adapted for multi-class predictions. When using the conventional approach, the GTs were binarized after preprocessing and data augmentation, the models were trained with a Dice loss, and sigmoid and softmax final activation functions were used for the single-class and multi-class models respectively.

An additional note about SofSeg: in the original work of SoftSeg, the final activation function used was a normalized ReLU. The ReLU prediction was then normalized by the maximum value to have a segmentation prediction corresponding to a level of confidence from 0 to 1. However, this activation function is not directly applicable to multi-class predictions as the different classes would not have probabilities summing up to 1. To generalize the normalized ReLU, the output of the original normalized ReLU is divided by the sum across all classes including the background class. This way, all predicted classes are mutually exclusive and have probabilities summing to 1.

\begin{table}[htb!]
\caption{Candidates’ description. The columns indicate the label fusion method while the rows present the training framework.}

\centering

\begin{tabular}{lccc}
\cmidrule[\heavyrulewidth]{2-4}

\multicolumn{1}{l}{\textbf{Training}} & \multicolumn{3}{c}{\textbf{Label fusion method}} \\
\cmidrule[\heavyrulewidth]{2-4}
\textbf{framework} & \textbf{STAPLE} & \textbf{Average}  & \textbf{Random sampling}\\
\midrule
\textbf{Conventional} & Conv-STAPLE & Conv-Average & Conv-RandomSamping \\
\textbf{SoftSeg} & SoftSeg-STAPLE & SoftSeg-Average & SoftSeg-RandomSamping \\
\bottomrule 
\end{tabular}

\label{tab:candidate}
\end{table}

\subsubsection{Training protocol}
All candidates were trained on 2D U-Net models. Training parameters for this work were the same as the one described in \cite{gros2021softseg} for the SCGM and MS brain lesion datasets. The use of the same model architecture, training parameters, dataset, and training environment (ivadomed) facilitate comparisons between the present study and the previous study by \citep{gros2021softseg}. Future studies could consider undertaking similar analyses with different model architectures, such as 3D models. The processing, training and evaluation pipeline is based on the open-source framework \url{ivadomed.org} \citep{gros2020ivadomed}. ivadomed is an open-source medical image analysis Python library based on PyTorch that provides tools, e.g., data loader, models, losses, transformations, pre- and post-processing, metrics, to train and use deep learning models for medical tasks such as segmentation.

\subsection{Datasets}
Two publicly available datasets with multiple raters were used to study label fusion: the SCGM challenge \citep{prados2017spinal} and MS brain lesion challenge \citep{commowick2018objective}.

\subsubsection{Gray and white matter challenge}
The SCGM dataset contains 80 T2*-weighted MRI of the cervical spinal cord, evenly acquired in four centers with different MR protocols and scanner vendors. Four raters segmented the gray and white matter from the scans using different guidelines and segmentation software which increases the inter-rater variability across centers. While the dataset includes 80 subjects, only 40 had all 4 raters publicly available, hence, this subdataset of 40 scans was retained for this study. A detailed description of the dataset and demographics of the scanned subjects and acquisition parameters can be found in \cite{prados2017spinal}  or a summary can be found in Appendix A.

\subsubsection{MS brain lesion challenge}
The MS brain lesion dataset containing MRI scans from 15 subjects was presented during the MICCAI 2016 challenge. MS lesions of each subject were annotated by seven expert raters. The dataset includes MRI scans with five contrasts: T1-weighted, T1-weighted Gadolinium-enhanced, T2-weighted, PD T2-weighted, and FLAIR. For a detailed description of the dataset see \cite{commowick2018objective}.

\subsection{Evaluation}
\subsubsection{Evaluation protocol}
Each model was trained with multiple random dataset splittings to limit splitting bias. For the SCGM dataset, 40 models were trained with an even split on the test centers (10 trainings with center 1 as test set, 10 trainings with center 2 as test set, etc.), while for the MS brain lesion dataset, 20 random splittings were performed (60/20/20\% for training/validation/test sets). Before assessing the predictions, the outputs were resampled in the native resolution. A non-parametric 2-sided Wilcoxon signed-rank test compared the most commonly used approach, “Conv-STAPLE”, with every other approach. Statistical difference was assessed by considering 0.05 as p-value threshold.

\subsubsection{Uncertainty due to inter-rater variability}
To evaluate the preservation of the inter-rater variability, we assessed the correspondence between the uncertainty from the prediction and the uncertainty associated with the multiple annotations. The patient uncertainty can be measured with the predictive entropy (Equation \ref{eq:entropy}) \citep{jungo2018effect} which can be directly compared with the entropy associated to the multiple rater segmentation (GT average). Entropy was chosen as it can directly be computed from the model’s prediction and does not require multiple forward passes as in other popular methods such as Monte Carlo dropout \citep{gal2016dropout} or test-test augmentation \citep{wang2019aleatoric}. A high entropy value indicates a high inter-rater variability. For example, if the fused label across raters is close to 0 or 1 in a given voxel, the level of agreement is high (i.e., low entropy), while values near 0.5 indicate high disagreement. A reliable model would generate a prediction reflecting the expert disagreement similarly to the GT average. Therefore, we plotted the entropy of the prediction against the entropy of the GT average and we expect the values to match. The correspondence was assessed by computing the mean absolute error (MAE) between both values for each patient data. 

\begin{equation}
\label{eq:entropy}
H = - \sum_{i=0}^{N_{vox}} p_i log(p_i)
\end{equation}

where $p_i$ is the model’s prediction for voxel $i$ and $N_{vox}$ is the total number of voxels in the image.

In addition, we quantified the voxel-wise similarity of the uncertain regions with voxels associated with high inter-rater variability. The Brier score (Equation \ref{eq:brier}) enables us to assess the similarity of non-binary data, hence was used to evaluate the similarity between the model’s prediction and the average labels from the expert raters. The average label from raters was selected as GT to quantify the performance of the soft prediction since information on inter-rater variability is encoded in this label while it cannot be directly observed from the STAPLE GT. The metric was computed for each segmentation class.

\begin{equation}
\label{eq:brier}
Brier score = \frac{1}{N_{vox}} \sum_{i=0}^{N_{vox}} (y_i - \hat{y}_i)^2
\end{equation}

where $y$ is the GT average,  $\hat{y}$ is the prediction, and $N_{vox}$ is the total number of voxels in the image.

\subsubsection{Calibration}
\label{sec:calibration}
Reliable deep learning models should predict calibrated outputs to truthfully indicate regions more prone to error or inter-expert disagreement. The model’s calibration quantifies how much the predicted values of a model truly represents the probability of the outcome, hence is an indicator of the quality of the model’s confidence. For instance, a perfectly calibrated model predicting 0.9 is confident at 90\% of its prediction and, therefore, should be correct 90\% of the time. Reliability diagrams \citep{degroot1983comparison} and the expected calibration error (ECE) \citep{naeini2015obtaining} were computed with the code from google-research repository\footnote{\url{https://github.com/google-research/google-research/blob/master/uncertainties/sources/postprocessing/metrics.py}} as used in \cite{guo2017calibration} to assess the calibration of the candidates.

\paragraph{Reliability diagram}
The reliability diagram helps to visualize the calibration of the model and plots the prediction’s accuracy (Equation \ref{eq:acc}) in relation to the model’s confidence (Equation \ref{eq:conf}). The identity function represents a perfectly calibrated model where the accuracy and the model’s confidence are always equal. Any deviation from this line can be translated into over- or underconfidence from the model. The model’s confidence was discretized into K=10 bins of size 0.1 ($\frac{1}{K}$) based on the literature \citep{naeini2015obtaining}. Moreover, preliminary results showed that changing the number of bins from K=5 to 20 did not change the calibration order of the approaches studied. We define confidence as the maximal prediction across classes for a given voxel. For a 3-class prediction problem, a model predicting [0.9, 0.06, 0.04] is associated with a confidence of 0.9. The minimum confidence for a 3-class prediction problem is $0.33^+$ (i.e., $[0.33^-, 0.33^-, 0.33^+]$), while for a binary prediction the minimum confidence is $0.5^+$ (i.e., $[0.5^-, 0.5^+]$). The predicted values are compared to the binarized GT, here, the STAPLE GT.  The accuracy is the proportion of voxels from a given bin, $B_k$, where the predicted class corresponds to the GT. The bin $B_m$ includes all predictions associated with a confidence of $[\frac{k}{K}, \frac{k+1}{K})$ \citep{guo2017calibration}. This accuracy is then compared to the average prediction in the bin. For instance, for the bin including voxels with values from [0.8 to 0.9), we expect that 85\% of the voxels in this bin, assuming uniform distribution of predicted values, are well classified. If the accuracy is greater than the model’s confidence, the model is underconfident, while a lower accuracy compared with the model’s confidence means the model is overconfident. 

\begin{equation}
\label{eq:acc}
Accuracy(B_k) = \frac{\sum_{i \in B_k} 1 (y_i = \hat{y}_i)}{\#B_k}
\end{equation}

\begin{equation}
\label{eq:conf}
Confidence(B_k) = \frac{\sum_{i \in B_k} \hat{y}_i}{\#B_k}
\end{equation}

where $\#B_k$ corresponds to the number of elements in the bin $B_k$.

\paragraph{Expected calibration error}
The reliability diagram does not display the information about the quantity of voxels in each bin. The ECE (Equation \ref{eq:ece}) is a metric extracted from the reliability diagram that takes into account the occurrence of voxels in each bin. The ECE corresponds to the sum of the absolute difference between the confidence of the model and the accuracy (i.e., the miscalibration) weighted by the number of voxels. The ECE was measured on all predictions from a model and averaged across models with different random splittings.

\begin{equation}
\label{eq:ece}
ECE = \sum_{k=0}^K \frac{\#B_k}{N_{vox}} \big{|} Accuracy(B_k) - Confidence(B_k)\big{|}
\end{equation}

where $N_{vox}$ is the total number of voxels in the image.

\subsubsection{Segmentation accuracy}
\paragraph{Metrics for binarized predictions}
To evaluate the quality of the segmentation the following metrics were used: Dice score, precision, recall, relative volume difference between the GT and prediction divided by the GT volume (RVD), and absolute volume difference (AVD) which is the absolute value of RVD. Due to the binary nature of these metrics, the predictions of the model were binarized. For example, a prediction of 0.5 with a GT of 0.5 obtained by averaging labels results in a Dice score of 0.5 even though both values are the same and should reach a maximal score. For this same reason, the STAPLE annotations were used as GTs for these metrics. Approaches using STAPLE during training (SoftSeg-STAPLE and Conv-STAPLE) were positively biased in that the evaluation metrics were computed with STAPLE as GT. Despite this limitation, STAPLE was the best option available for binary GT as it takes into account all the raters’ opinions at once. For the MS dataset which has two classes (i.e., lesion or background), the binarization threshold was found by searching for the optimal value (between 0 and 1 with an increment of 0.05) in terms of Dice score as done in \cite{gros2021softseg}. The threshold was optimized for each model individually, regardless of the training method (i.e., conventional or SoftSeg models). For the SCGM dataset, there are three classes: gray matter, white matter and background. The predicted class is obtained by selecting the maximum prediction across the three classes.

\paragraph{Composite score}
To represent the overall segmentation accuracy performance, a composite score is computed by aggregating the above metrics. Firstly, z-scores for each metric are derived by standardizing the results across candidates (i.e., zero mean and unit standard deviation). Secondly, the z-scores are linearly aggregated to compute the composite score, with equal absolute weights across metrics. A weight of 1 was used for the Dice, precision, and recall (because they need to be maximized), and a weight of -1 was used for the AVD (because it needs to be minimized).

\section{Results}

\subsection{Inter-rater uncertainty}

\begin{figure}[H]
\begin{center}
    \subfloat[\centering SCGM]{\includegraphics[width=0.8\linewidth]{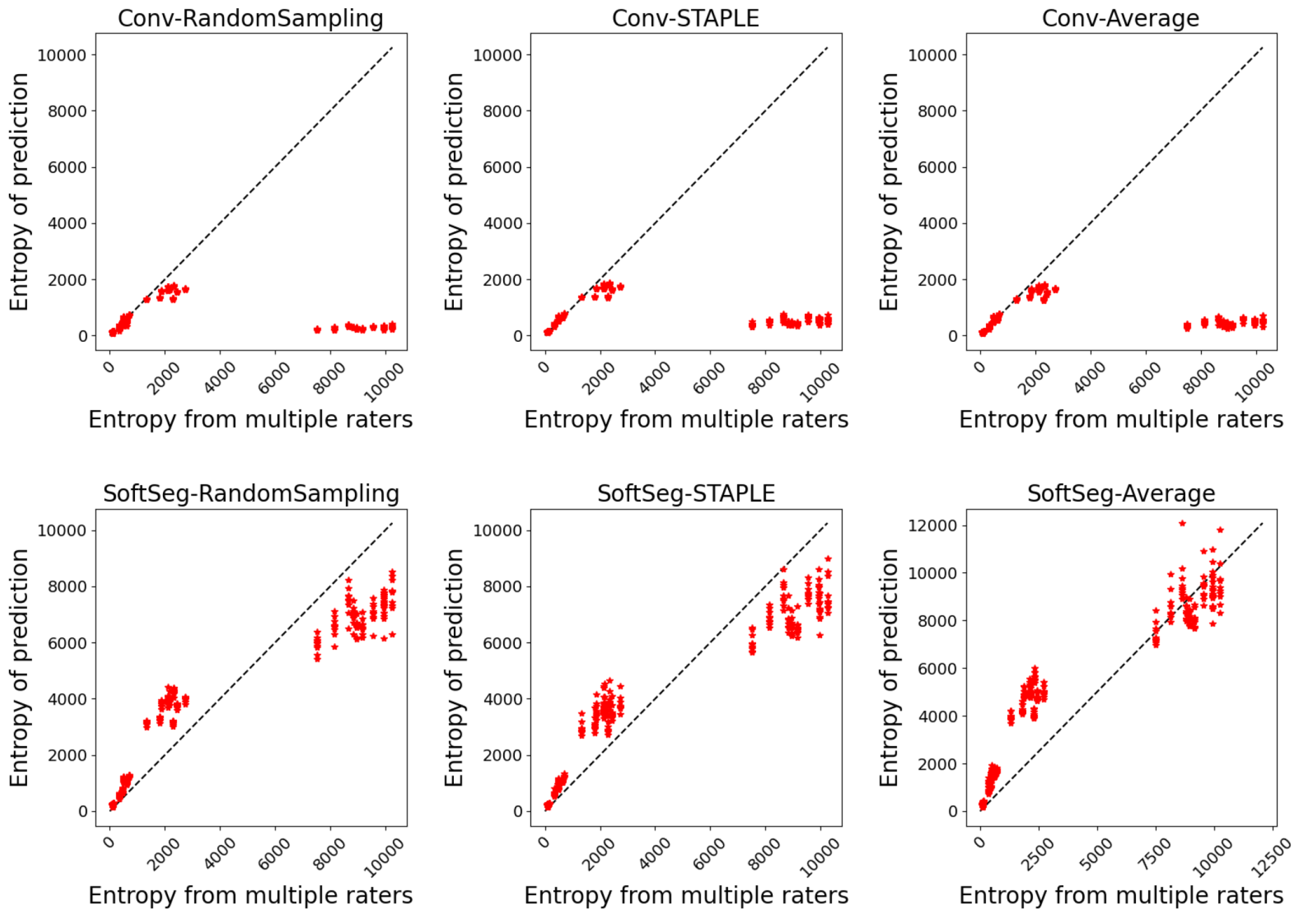} \label{fig:unc_scgm}}%
    \qquad
    \subfloat[\centering MS brain]{\includegraphics[width=0.8\linewidth]{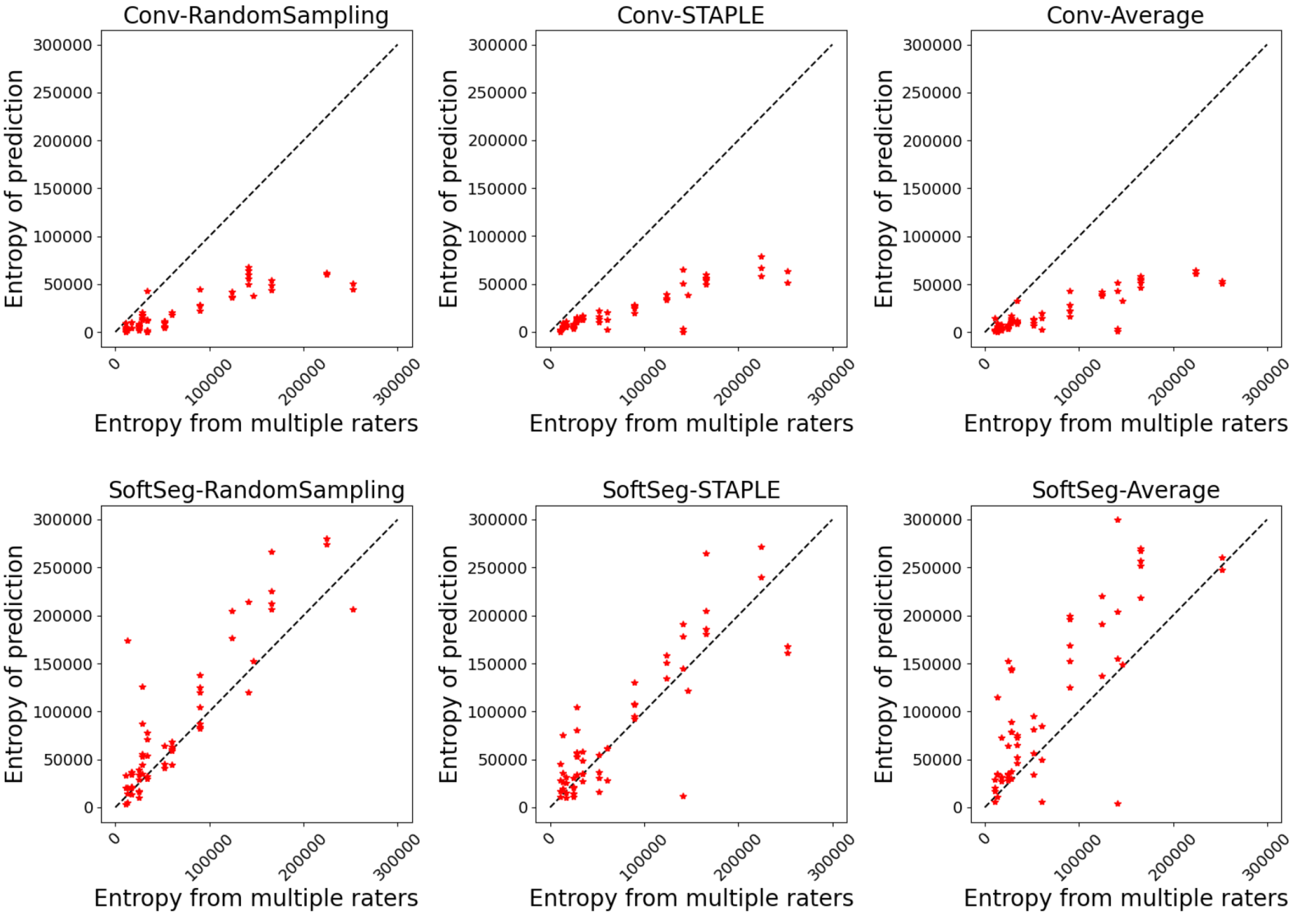} \label{fig:unc_ms}}
\end{center}
\caption{Comparison of entropy generated by inter-rater variability and entropy from the model's prediction for the SCGM (a) and MS brain lesions (b) datasets. Each red dot corresponds to a participant. The dashed line represents the identity line where data points from an ideal model should lie.}
\label{fig:unc}
\end{figure}

The predicted segmentation should ideally reflect the uncertainty associated with the disagreement between experts. Figure \ref{fig:unc} illustrates the agreement between the entropy generated from the multiple expert ratings and the predicted segmentation’s entropy. Similar observations can be drawn for both SCGM and MS brain segmentation. The SoftSeg models showed better correspondence between the predicted and true entropy which can be seen by data points lying near the identity line (perfect agreement) and smaller MSE values. All models trained with the conventional framework showed a tendency to generate less entropy which can be interpreted as overconfidence and an underestimation of the uncertainty. Random sampling and STAPLE have similar patterns with the SoftSeg framework and reflect the more truthfully the entropy linked to multiple raters. “SoftSeg-Average” showed a slight tendency to overestimate the uncertainty. Clusters can be observed in Figure \ref{fig:unc_scgm} and are associated with the different data centers of the SCGM dataset.

Table \ref{tab:unc} summarizes the metrics associated with the preservation of the inter-rater variability. A general trend that can be observed is that SoftSeg candidates performed better than their conventional counterparts. When performing pairwise comparisons of each candidate using SoftSeg vs. the conventional framework, SoftSeg systematically yielded the best average metric. More precisely, for all metrics, “SoftSeg-RandomSampling” and “SoftSeg-STAPLE” were always the top two performing candidates. For both dataset and on all classes, “SoftSeg-RandomSampling” yielded the lowest Brier score indicating the greater resemblance with the segmentation from the averaged labels. “SoftSeg-STAPLE” obtained the best correspondence, i.e., lowest MAE, between the predicted uncertainty and the inter-rater variability. The MAE, which should be minimized, of conventional models was on average 46\% and 45\% higher compared with SoftSeg models for the SCGM and MS brain datasets respectively. 

\newcommand{\STAB}[1]{\begin{tabular}{@{}c@{}}#1\end{tabular}}

\begin{table}
\caption{Quantitative assessment of the inter-rater variability preservation on the SCGM and MS brain datasets (MEAN ± STD). Brier score is reported by segmentation class while the MAE is computed on the total entropy of the entire image. Each row represents a candidate. The best averaged result for each metric and tissue is displayed in bold. Statistical differences are computed between “Conv-STAPLE” (ref) and each other candidate (**: $p<0.05$). Abbreviations: Opt.: optimal;  MAE: mean absolute error; GM: gray matter; WM: white matter.}

\resizebox{\textwidth}{!}{

\begin{tabular}{lcccc|cc}
\cmidrule[\heavyrulewidth]{3-7}
\multicolumn{2}{l}{} & \multicolumn{3}{c}{\textbf{SCGM}} & \multicolumn{2}{c}{\textbf{MS brain}} \\
\cmidrule[\heavyrulewidth]{3-7}
 & & \multicolumn{2}{c}{\thead{\textbf{Brier score} ($\times 10^{-4}$) \\ \textit{Opt. value: 0}}} & \thead{\textbf{MAE} ($\times 10^3$) \\ \textit{Opt. value: 0}}  & \thead{\textbf{Brier score} ($\times 10^{-4}$) \\ \textit{Opt. value: 0}} & \thead{\textbf{MAE} ($\times 10^4$) \\ \textit{Opt. value: 0}}\\
\cmidrule[\heavyrulewidth]{3-7}
 & & \textbf{GM} & \textbf{WM} & \textbf{Entire image} & \textbf{MS lesions} & \textbf{Entire image} \\
 \toprule
 \multirow{3}{*}{\rotatebox[origin=c]{90}{\footnotesize{\STAB{\textbf{Conventional}}}}} & \textbf{\thead{STAPLE \\ (ref)}} & $1.05 \pm 0.04$ & $3.08 \pm 0.11$ & $2.30 \pm 0.02$ & $7.78 \pm 5.84$ & $5.32 \pm 2.86$ \\
 & \textbf{Average} & $1.18 \pm 0.07*$ & $3.02 \pm 0.11*$ & $2.33 \pm 0.01*$ & $7.91 \pm 5.05$ & $5.43 \pm 3.07$ \\
& \textbf{\thead{Random \\ sampling}} & $0.98 \pm 0.05*$ & $2.81 \pm 0.16*$ & $2.37 \pm 0.01*$ & $8.21 \pm 3.84$ & $4.86 \pm 2.53$ \\
\midrule
\multirow{3}{*}{\rotatebox[origin=c]{90}{\STAB{\textbf{SoftSeg}}}} & \textbf{STAPLE} & $0.96 \pm 0.06*$ & $2.92 \pm 0.10*$ & $\bm{0.98 \pm 0.06}^*$ & $5.71 \pm 2.74*$ & $\bm{2.97 \pm 2.59}^*$ \\
& \textbf{Average} & $1.14 \pm 0.07*$ & $3.00 \pm 0.10*$ & $1.13 \pm 0.09*$ & $5.55 \pm 2.91*$ & $5.19 \pm 3.20$ \\
& \textbf{\thead{Random \\ sampling}} & $\bm{0.90 \pm 0.03}^*$ & $\bm{2.68 \pm 0.03}^*$ & $1.08 \pm 0.08^*$ & $\bm{5.01 \pm 2.41}^*$ & $3.40 \pm 3.96^*$ \\
\bottomrule 
\end{tabular}}

\label{tab:unc}
\end{table}

\subsection{Visual assessment}
\begin{figure}[]
\centering
\includegraphics[width=0.55\textwidth]{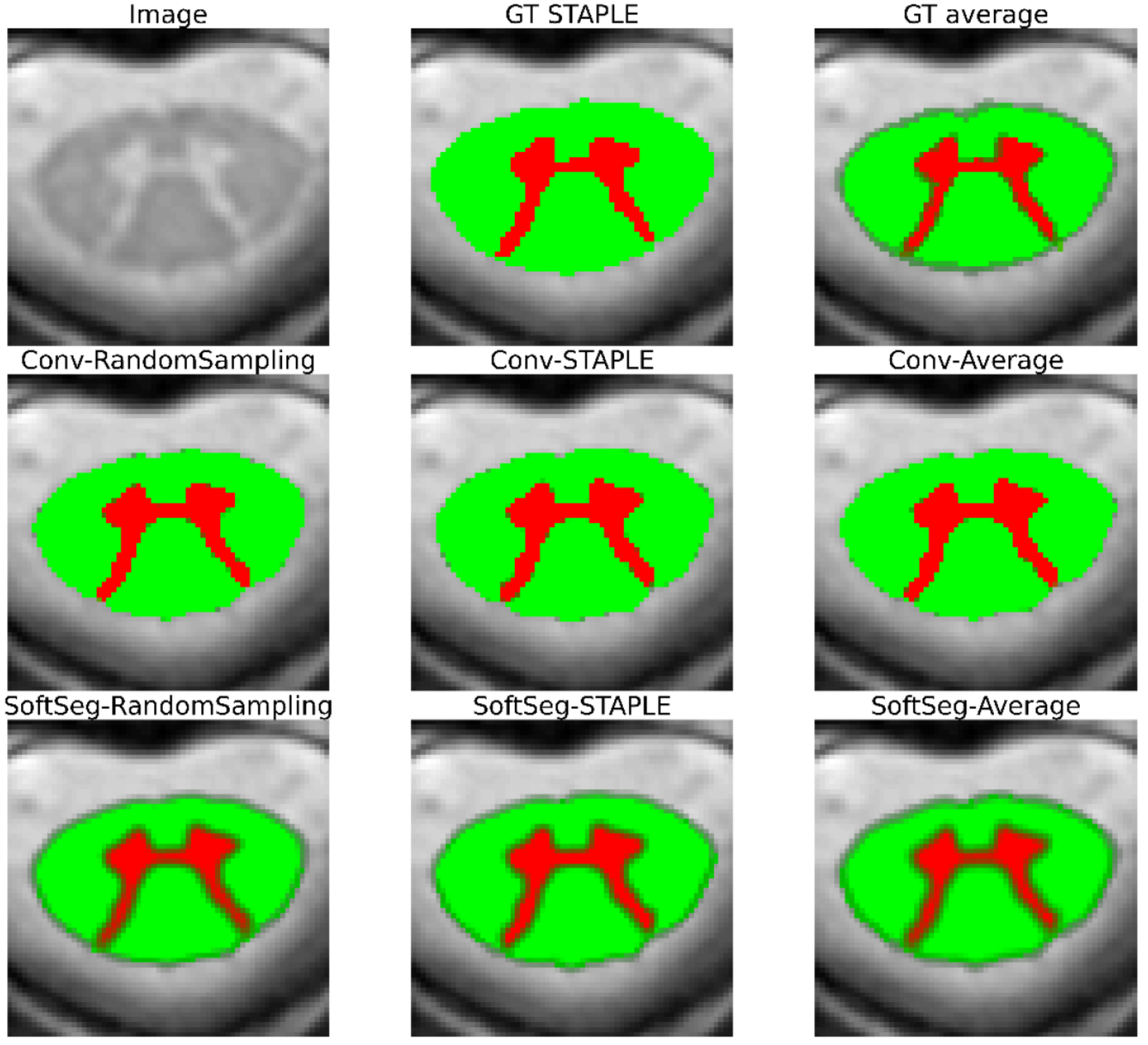}
\caption{Visual assessment of STAPLE and average GTs and predictions from the six candidates on spinal gray and white matter segmentation. Abbreviations: GT: ground truth.}
\label{fig:viz_scgm}
\end{figure}

\begin{figure}[]
\centering
\includegraphics[width=0.65\textwidth]{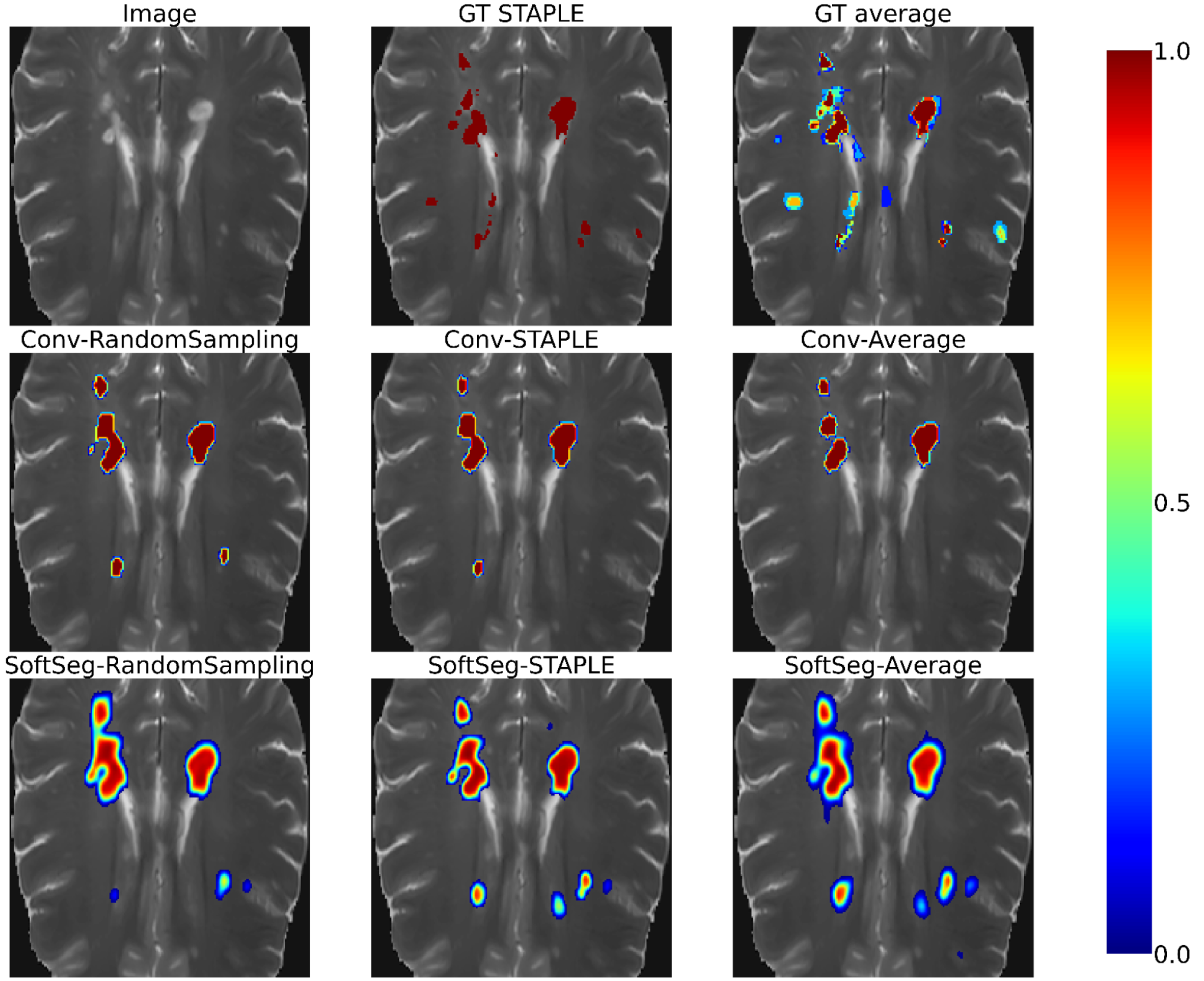}
\caption{Visual assessment of STAPLE and GT average and predictions from the six candidates on MS brain segmentation. Red: Gray matter. Green: White matter. Voxels at tissue boundaries represent values between 0 and 1. Abbreviations: GT: ground truth.}
\label{fig:viz_ms}
\end{figure}

Figure \ref{fig:viz_scgm} and Figure \ref{fig:viz_ms} contain the segmentations from the STAPLE and GT average and from the predictions of the six candidates. Regardless of the label fusion method and the dataset, predictions using conventional models have sharp edges between tissue types (similar to the GT STAPLE) and underestimated the inter-rater variability. All SoftSeg candidates display smoother boundaries (similar to the GT average). When comparing the SoftSeg models, “SoftSeg-Average” presents the softest edges followed by “SoftSeg-RandomSampling”, then “SoftSeg-STAPLE”. These differences are especially noticeable in Figure \ref{fig:viz_scgm} at the extremity of the dorsal horns and near the central canal (black arrows) and in Figure \ref{fig:viz_ms} on the lesion aggregate on the top-left (yellow arrows). An ideal prediction should reflect the inter-rater variability similarly to the GT average. Hence, predictions should not be too sharp or too smooth compared to the GT average.

\subsection{Calibration}
\begin{figure}[H]
\begin{center}
    \subfloat[\centering SCGM]{\includegraphics[width=0.9\linewidth]{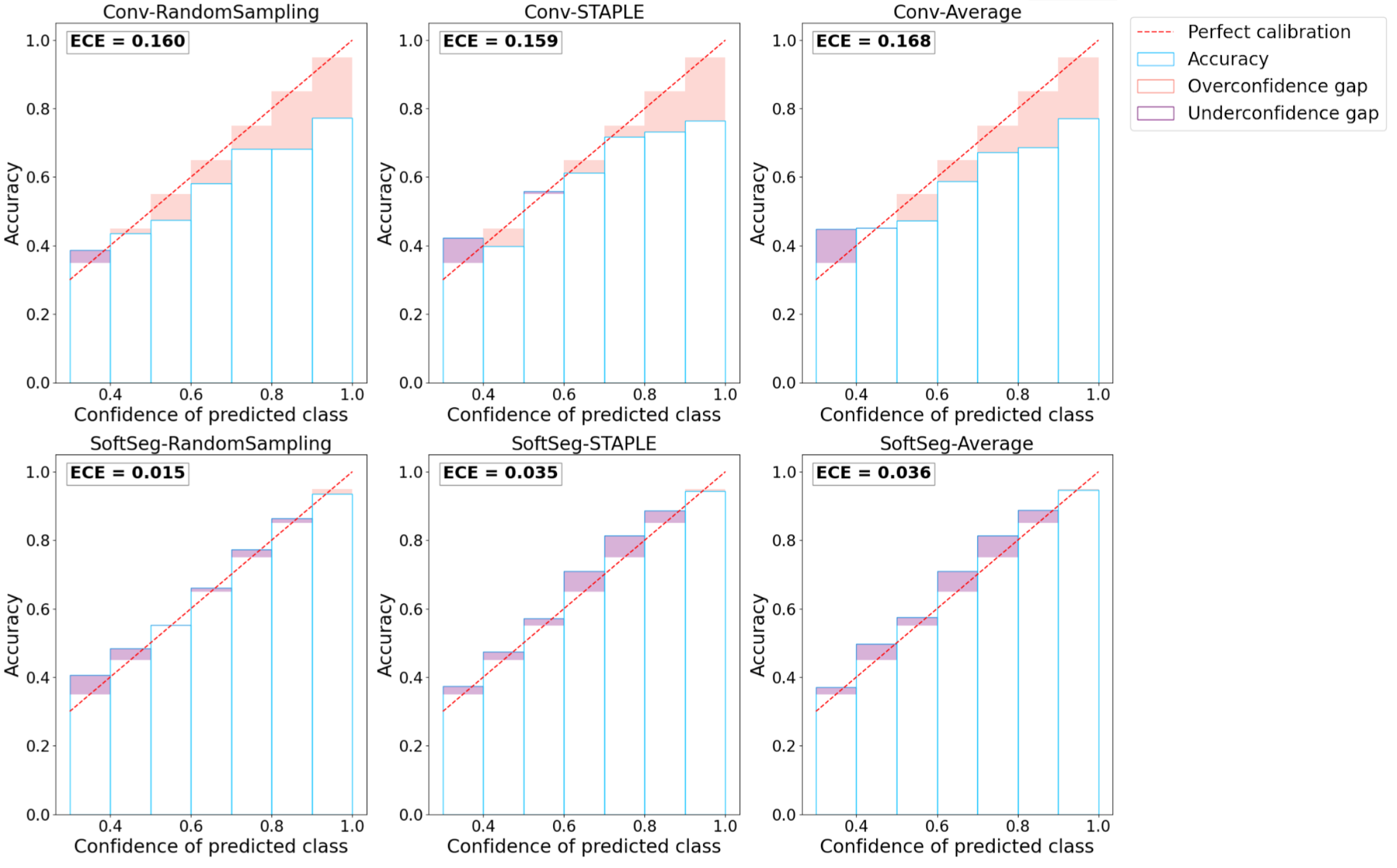} \label{fig:cal_scgm}}%
    \qquad
    \subfloat[\centering MS brain]{\includegraphics[width=0.9\linewidth]{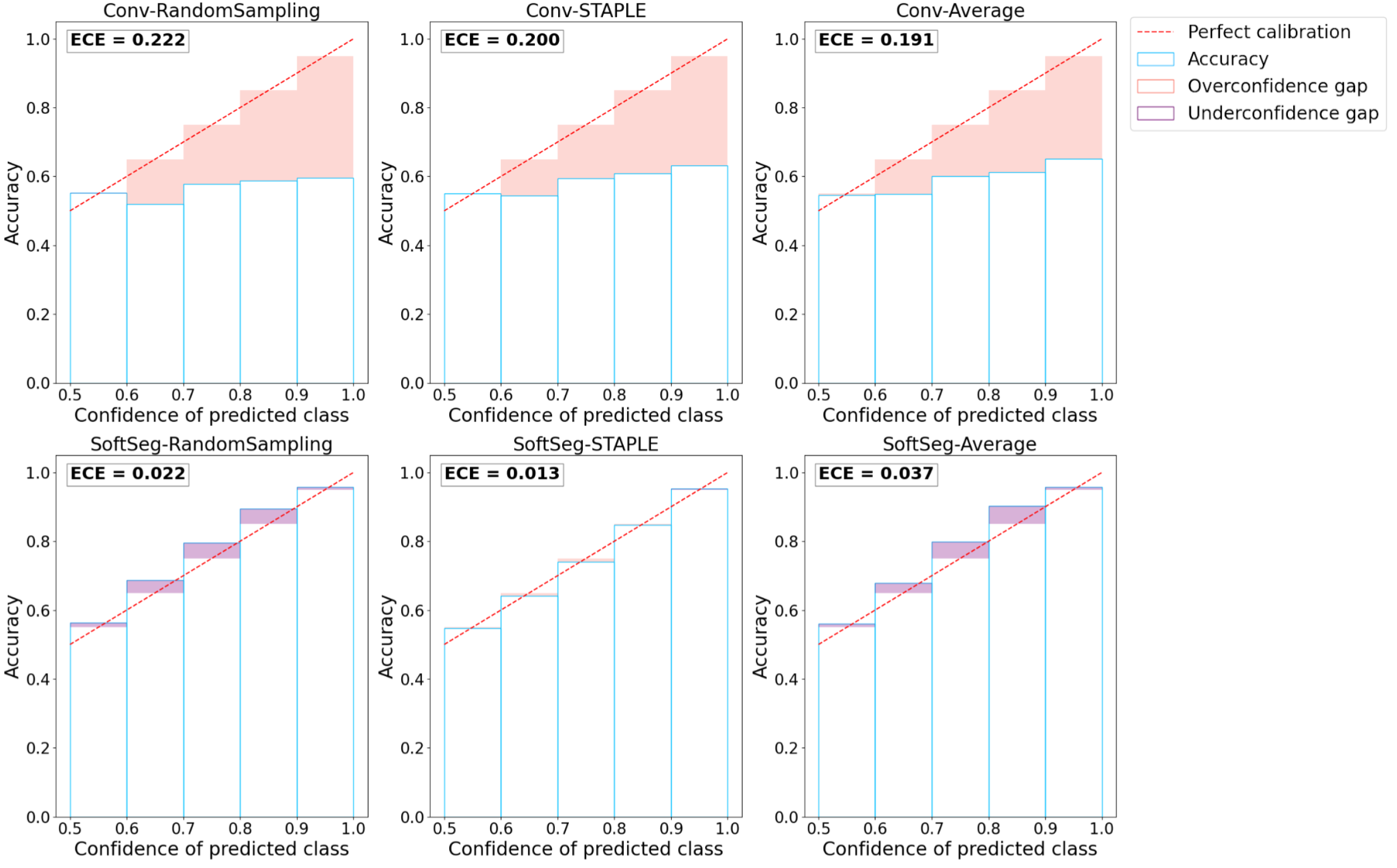} \label{fig:cal_ms}}
\end{center}
\caption{Reliability diagram for all candidates on SCGM (a) and MS brain lesions (b) datasets. The red identity line illustrates a perfect calibration. Orange bands represent overconfidence while the purple ones indicate underconfidence.}
\label{fig:cal}
\end{figure}

Figure \ref{fig:cal} presents the reliability diagrams generated from the predictions on SCGM and on MS brain lesion datasets. The conventional approach is overconfident with most of its predictions for all the datasets and label fusion methods. This overconfidence results in high ECE: 16.2\% for SCGM and 20.4\% for MS lesions on average. In contrast, the SoftSeg is on average better calibrated with an ECE of 2.9\% for SCGM and 2.4\% for MS lesions. SoftSeg candidates mostly present slight underconfidence with the exception of “SoftSeg-STAPLE” on MS lesions which is overall well calibrated with minimal overconfidence. “SoftSeg-STAPLE” and “SoftSeg-RandomSampling” are the candidates presenting the best calibration. “SoftSeg-Average” presents more underconfidence compared to the other SoftSeg candidates due to the overly soft predictions encouraged by the non-binary GT.

\subsection{Segmentation accuracy}
\begin{figure}[H]
\begin{center}
    \subfloat[\centering SCGM]{\includegraphics[width=0.7\linewidth]{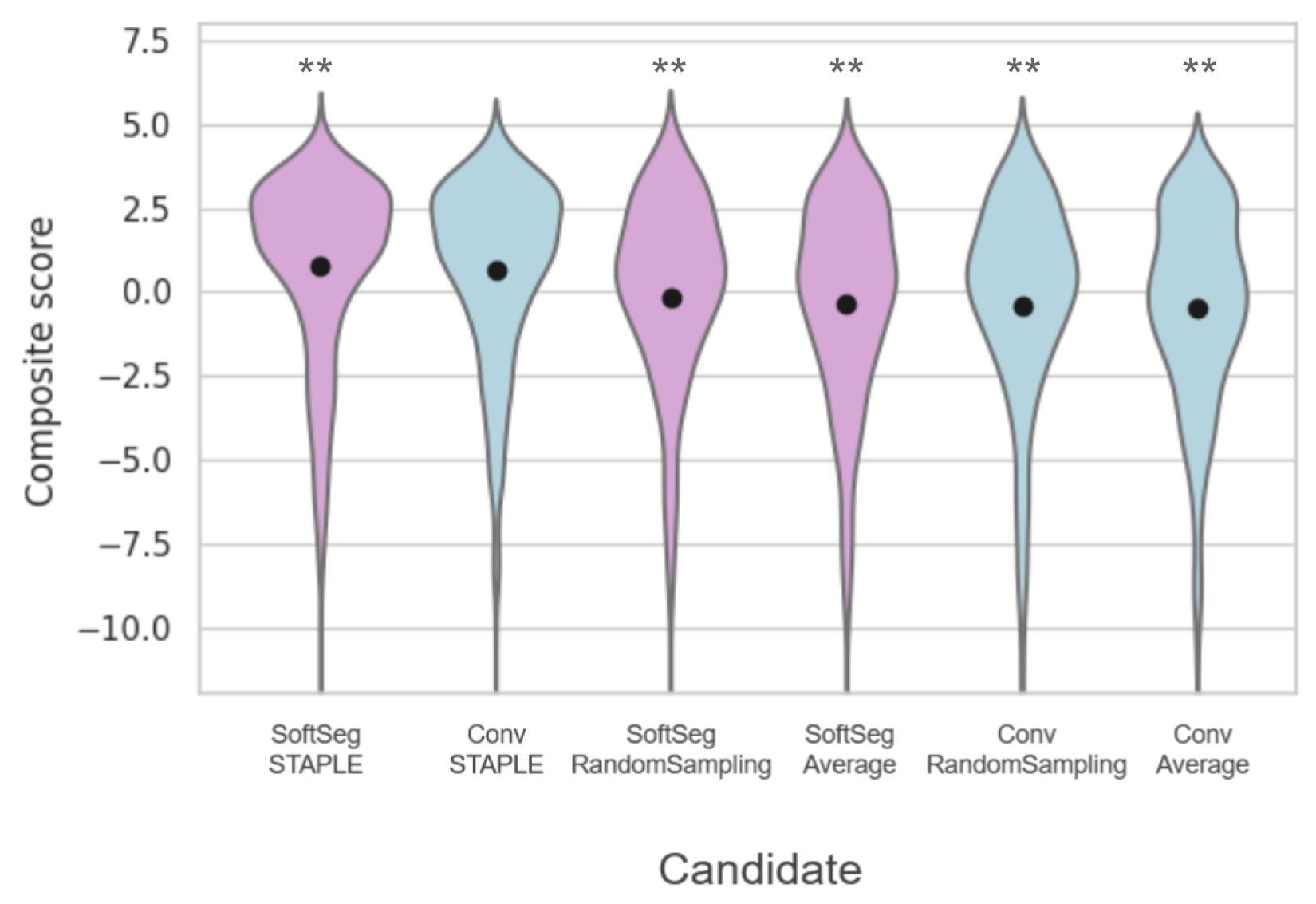} \label{fig:comp_scgm}}%
    \qquad
    \subfloat[\centering MS brain]{\includegraphics[width=0.7\linewidth]{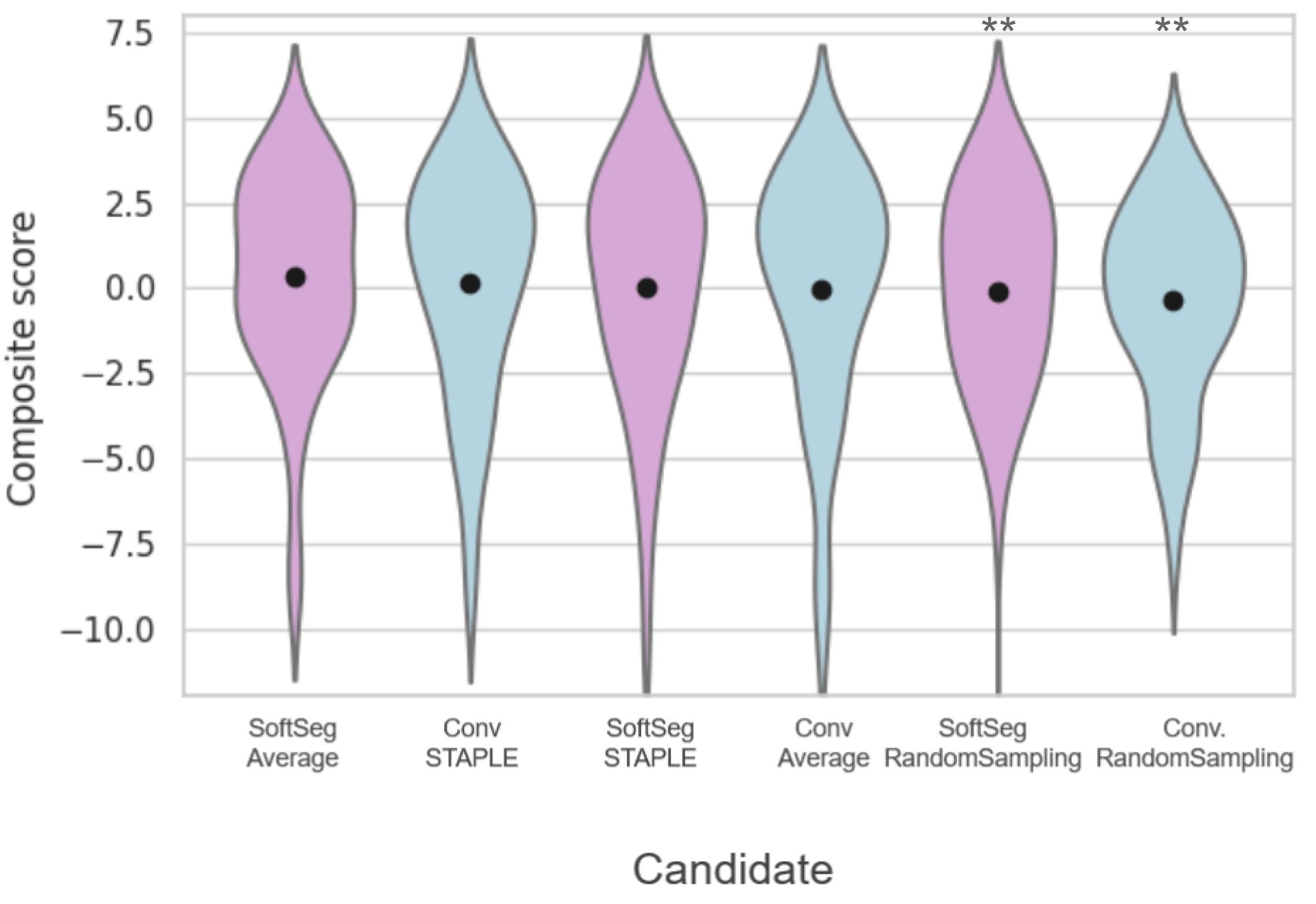} \label{fig:comp_ms}}
\end{center}
\caption{Composite score across candidates on the SCGM (a) and MS brain (b) datasets. The composite score aggregates the following segmentation metrics: Dice, Precision, Recall, and AVD. For each candidate, each violin plot represents the distribution of composite scores across testing patients and random splittings. They are sorted from the best to the worst averaged composite score (black dot). Statistical differences are computed between “Conv-STAPLE” (ref) and each other candidate (**: $p<0.05$).}
\label{fig:composite}
\end{figure}

Table \ref{tab:seg_scgm} presents the quantitative results of the segmentation accuracy assessment on the SCGM dataset. When comparing the binarized predictions to the STAPLE GT, “SoftSeg-STAPLE” yielded the best Dice, recall, AVD, and RVD score for both white and gray matter segmentation and significantly outperformed the “Conv-STAPLE” method ($p<0.05$). Figure \ref{fig:comp_scgm} summarizes the metrics presented in Table 3a using a composite score. The averaged composite score indicates that, regardless of the training pipeline (i.e., conventional and SoftSeg), the best label fusion method is STAPLE (see Figure \ref{fig:comp_scgm}). The composite score of SoftSeg was consistently higher compared with the conventional framework for a given label fusion method. All composite scores were statistically different from the “Conv-STAPLE” candidate.

Table \ref{tab:seg_ms} introduces the MS brain segmentation performance metrics. Most metrics on the binary prediction demonstrated no significant difference compared with the “Conv-STAPLE” candidate. Only the “Conv-RandomSampling” candidate had a significantly lower Dice score compared to the “Conv-STAPLE”. Figure \ref{fig:comp_ms} summarizes the metrics presented in Table \ref{tab:seg_ms} using a composite score. “SoftSeg-Average” provided the best composite score, followed by “Conv-STAPLE”. When comparing the composite scores of the candidates with “Conv-STAPLE”, no significant difference was found, except “Conv-RandomSampling” and “SoftSeg-RandomSampling” which led to significantly lower results.

\begin{table}[H]
\caption{Quantitative assessment of the segmentation performance on the SCGM and brain MS lesions datasets. For SCGM, each value represents the average over 40 models and intervals are the standard deviation over these. Mean and standard deviation are reported for both gray and white matters. For brain MS lesions, each value represents the average and standard deviation over 20 models and intervals are the standard deviation over these. Each row represents a candidate. The best averaged result for each metric and tissue is displayed in bold. All metrics are computed on binarized predictions against the “STAPLE GT”. Statistical differences are computed between “Conv-STAPLE” (ref) and each other candidate (*: $p<0.05$). Abbreviations: Opt.: optimal; GM: gray matter; WM: white matter.}

\begin{subtable}{\textwidth}
\caption{SCGM}
 \resizebox{\textwidth}{!}{
\begin{tabular}{lccccccccccc}
\cmidrule[\heavyrulewidth]{3-12}
 & & \multicolumn{2}{c}{\thead{\textbf{Dice [\%]} \\ \textit{Opt. value: 100}}} & \multicolumn{2}{c}{\thead{\textbf{Precision [\%]} \\ \textit{Opt. value: 100}}}  & \multicolumn{2}{c}{\thead{\textbf{Recall [\%]} \\ \textit{Opt. value: 100}}} & \multicolumn{2}{c}{\thead{\textbf{AVD [\%]} \\ \textit{Opt. value: 0}}} & \multicolumn{2}{c}{\thead{\textbf{RVD [\%]} \\ \textit{Opt. value: 0}}}\\
 \cmidrule[\heavyrulewidth]{3-12}
 && \textbf{GM} & \textbf{WM} & \textbf{GM} & \textbf{WM}  & \textbf{GM} & \textbf{WM} & \textbf{GM} & \textbf{WM} & \textbf{GM} & \textbf{WM} \\
 \toprule
 \multirow{3}{*}{\rotatebox[origin=c]{90}{\footnotesize{\STAB{\textbf{Conventional}}}}} & \textbf{\thead{STAPLE \\ (ref)}} & \thead{$86.6$ \\ $\pm 2.8$} & \thead{$93.1$ \\ $\pm 2.0$} & \thead{$89.0$ \\ $\pm 6.5$} & \thead{$94.0$ \\ $\pm 1.0$} & \thead{$85.1$ \\ $\pm 3.1$} & \thead{$92.5$ \\ $\pm 3.5$} & \thead{$10.1$ \\ $\pm 3.8$} & \thead{$\bm{5.2}$ \\ $\bm{\pm 1.8}$} & \thead{$3.6$ \\ $\pm 9.9$} & \thead{$1.5$ \\ $\pm 3.5$} \\
 & \textbf{Average} & \thead{$83.6$ \\ $\pm 2.1^*$} & \thead{$90.7$ \\ $\pm 2.6^*$} & \thead{$91.6$ \\ $\pm 6.7^*$} & \thead{$94.5$ \\ $\pm 0.7^*$} & \thead{$77.9$ \\ $\pm 5.1^*$} & \thead{$87.5$ \\ $\pm 4.8^*$} & \thead{$17.4$ \\ $\pm 6.9^*$} & \thead{$7.9$ \\ $\pm 4.6^*$} & \thead{$14.1$ \\ $\pm 12.1^*$} & \thead{$7.4$ \\ $\pm 5.1^*$} \\
& \textbf{\thead{Random \\ sampling}} & \thead{$84.2$ \\ $\pm 4.4^*$} & \thead{$90.1$ \\ $\pm 4.2^*$} & \thead{$\bm{93.6}$ \\ $\bm{\pm 5.4^*}$} & \thead{$\bm{95.5}$ \\ $\bm{\pm 0.4^*}$} & \thead{$77.8$ \\ $\pm 7.4^*$} & \thead{$86.0$ \\ $\pm 6.6^*$} & \thead{$18.1$ \\ $\pm 9.4^*$} & \thead{$10.1$ \\ $\pm 6.7^*$} & \thead{$16.3$ \\ $\pm 11.9^*$} & \thead{$10.0$ \\ $\pm 6.8^*$} \\
\midrule
\multirow{3}{*}{\rotatebox[origin=c]{90}{\STAB{\textbf{SoftSeg}}}} & \textbf{STAPLE} & \thead{$\bm{87.1}$ \\ $\bm{\pm 3.4^*}$} & \thead{$\bm{93.3}$ \\ $\bm{\pm 2.1^*}$} & \thead{$88.7$ \\ $\pm 7.3^*$} & \thead{$93.7$ \\ $\pm 1.0^*$} & \thead{$\bm{86.4}$ \\ $\bm{\pm 2.5^*}$} & \thead{$\bm{93.2}$ \\ $\bm{\pm 3.7^*}$} & \thead{$\bm{9.9}$ \\ $\bm{\pm 4.5^*}$} & \thead{$\bm{5.2}$ \\ $\bm{\pm 2.0^*}$} & \thead{$\bm{1.6}$ \\ $\bm{\pm 10.7^*}$} & \thead{$\bm{0.5}$ \\ $\bm{\pm 3.7^*}$} \\
& \textbf{Average} & \thead{$83.7$ \\ $\pm 3.6^*$} & \thead{$91.0$ \\ $\pm 2.5^*$} & \thead{$91.1$ \\ $\pm 7.5^*$} & \thead{$95.0$ \\ $\pm 0.8^*$} & \thead{$78.8$ \\ $\pm 6.1^*$} & \thead{$87.5$ \\ $\pm 4.4^*$} & \thead{$16.8$ \\ $\pm 7.5^*$} & \thead{$8.2$ \\ $\pm 4.5^*$} & \thead{$12.4$ \\ $\pm 13.6^*$} & \thead{$7.9$ \\ $\pm 4.6^*$} \\
& \textbf{\thead{Random \\ sampling}} & \thead{$84.8$ \\ $\pm 4.0$} & \thead{$90.4$ \\ $\pm 3.6^*$} & \thead{$93.1$ \\ $\pm 5.6^*$} & \thead{$\bm{95.5}$ \\ $\bm{\pm 0.8^*}$} & \thead{$79.1$ \\ $\pm 6.3^*$} & \thead{$86.6$ \\ $\pm 5.8^*$} & \thead{$16.4$ \\ $\pm 8.1^*$} & \thead{$9.4$ \\ $\pm 6.0^*$} & \thead{$14.4$ \\ $\pm 11.0^*$} & \thead{$9.2$ \\ $\pm 6.2^*$} \\
\bottomrule 
\end{tabular}
}

\label{tab:seg_scgm}%
\end{subtable}

\begin{subtable}{\textwidth}
\caption{MS brain}
 \resizebox{\textwidth}{!}{

\begin{tabular}{lcccccc}
\cmidrule[\heavyrulewidth]{3-7}
 & & \thead{\textbf{Dice [\%]} \\ \textit{Opt. value: 100}} & \thead{\textbf{Precision [\%]} \\ \textit{Opt. value: 100}}  & \thead{\textbf{Recall [\%]} \\ \textit{Opt. value: 100}} & \thead{\textbf{AVD [\%]} \\ \textit{Opt. value: 0}} & \thead{\textbf{RVD [\%]} \\ \textit{Opt. value: 0}}\\
 \toprule
 \multirow{3}{*}{\rotatebox[origin=c]{90}{\footnotesize{\STAB{\textbf{Conventional}}}}} & \textbf{\thead{STAPLE \\ (ref)}} & $54.9 \pm 12.8$ & $56.3 \pm 14.0$ & $\bm{57.0 \pm 12.4}$ & $31.8 \pm 18.2$ & $-7.5 \pm 26.0$ \\
 & \textbf{Average} & $54.3 \pm 12.8$ & $58.0 \pm 14.8$ & $55.6 \pm 13.0^*$ & $42.0 \pm 57.1$ & $-11.8 \pm 65.5$ \\
& \textbf{\thead{Random \\ sampling}} & $50.6 \pm 11.8^*$ & $61.4 \pm 13.4$ & $52.9 \pm 12.6$ & $54.8 \pm 54.8^*$ & $-9.9 \pm 64.3$ \\
\midrule
\multirow{3}{*}{\rotatebox[origin=c]{90}{\STAB{\textbf{SoftSeg}}}} & \textbf{STAPLE} & $55.0 \pm 11.1$ & $\bm{60.4 \pm 13.2}$ & $56.6 \pm 11.1$ & $51.2 \pm 56.2$ & $-16.6 \pm 63.2$ \\
& \textbf{Average} & $\bm{56.0 \pm 10.7}$ & $59.8 \pm 13.0$ & $55.5 \pm 11.5$ & $\bm{30.6 \pm 17.7}$ & $\bm{-1.7 \pm 26.0}$ \\
& \textbf{\thead{Random \\ sampling}} & $53.7 \pm 10.3$ & $58.5 \pm 13.1$ & $52.8 \pm 9.4^*$ & $39.7 \pm 50.2$ & $-8.3 \pm 59.2$ \\
\bottomrule 
\end{tabular}
}
    \label{tab:seg_ms}
\end{subtable}
\label{tab:seg}
\end{table}

\section{Comparison with loss ensembles}
\label{sec:ss-loss-ensembles}

In this section, we compare our method to loss ensembles \cite{ma2020estimating}  - winner of the QUBIQ 2020 challenge. Loss ensembles are inspired by the observation that the manual labels are typically generated by several expert radiologists with varying expertise. Therefore, during the training phase, this method requires training of multiple independent models, each with labels corresponding to a different rater. During inference, the test input was passed through each model and the output was averaged. 

\begin{figure}[htbp!]
\centering
\includegraphics[width=0.9\linewidth]{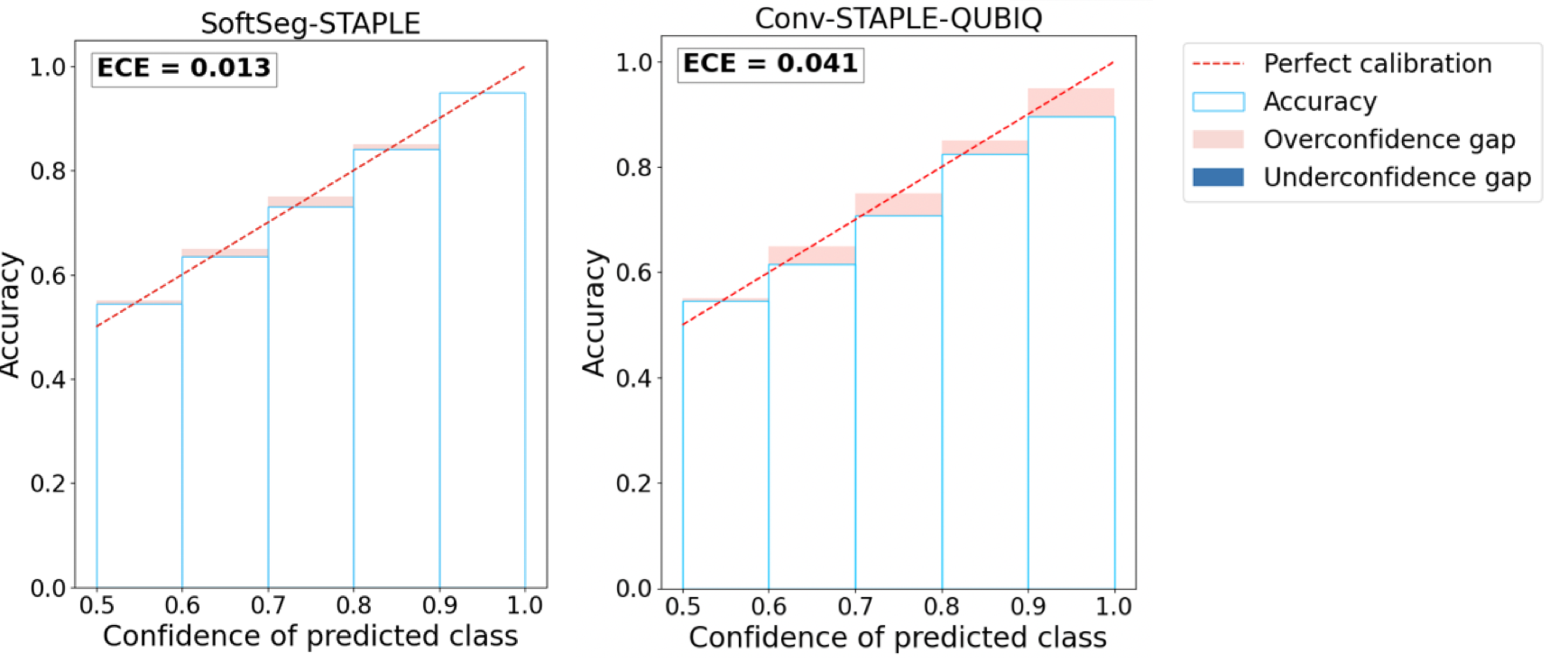} 
\caption{Reliability diagrams comparing our method SoftSeg-STAPLE (left) with the Deep Ensembles method (right). 
The red identity line illustrates a perfect calibration and the orrange bands represent overconfidence.}
\label{fig:ss-qubic}
\end{figure}

We used the MS brain lesion dataset \cite{commowick2016msseg} for this experiment, containing GT labels from 7 expert raters. Therefore, we trained 7 separated 2D UNet models, each corresponding to a different rater. The final test output was the averaged across the predictions for all the raters. We call this method "Conv-STAPLE-QUBIQ". Figure \ref{fig:ss-qubic} shows the reliability diagrams for our method and the loss ensembles. We observed that our method SoftSeg-STAPLE is relatively less over-confident compared to loss ensembles \cite{ma2020estimating} and also achieves lower ECE. 

We also computed the Brier Score and the MAE of these two methods to better quantify the uncertainty arising from inter-rater variability (see Table \ref{tab:mae_brier_qubic}). It can be seen that SoftSeg-STAPLE performs better on both the uncertainty quantification metrics (while also being computationally efficient).

\begin{table}[htbp!]
\centering
\caption{Comparison based on Inter-rater Uncertainty Quantification Metrics (lower the better)}
\begin{tabular}{lcc}    
\toprule
{\textbf{Method}} & \textbf{Brier Score $( \times 10^{-4})$} & \textbf{MAE $(\times 10^{4})$} \\ 
\midrule
    Loss Ensembles \cite{ma2020estimating} &  $5.39 \pm 1.94$   &   $1.08 \pm 0.59$  \\
    SoftSeg-STAPLE & \textbf{5.28 $\pm$ 1.83}  &    \textbf{1.01 $\pm$ 0.50}    \\
\bottomrule
\end{tabular}
\label{tab:mae_brier_qubic}
\end{table}

\section{Discussion}
Data labeling is prone to inter-rater variability, and it is still unclear how to best preserve this valuable information when training a deep learning model. In this study, we compared three label fusion methods, using both SoftSeg or a conventional training framework. Overall, SoftSeg models were shown to provide a more reliable representation of the inter-rater variability than using the conventional approach, in terms of (i) correspondence between the predicted and true uncertainty, (ii) visual assessment, (iii) calibration, and (iv) segmentation accuracy. This study suggests that the conventional framework has a tendency to be overconfident and to underestimate the uncertainty, regardless of the label fusion method used. When using SoftSeg, random sampling and STAPLE label fusion methods showed a more reliable inter-rater uncertainty and calibration than the average label fusion method. In this section, we further discuss avenues to preserve information from all raters, via label fusion and/or training pipeline, then we discuss the need to go beyond the Dice score for model evaluations, particularly in the context of multiple raters and soft predictions. Finally, we discuss the importance of repeatability in medical deep learning research.

\subsection{The preservation of the inter-rater variability}
Encoding the inter-rater variability in the model training is important as it helps tailor models that reflect the experts’ disagreement through the predictions. In this study, we investigated two avenues to preserve the inter-rater variability when training a deep learning model: how the raters’ labels are fused, and how the labels are processed by the training framework. Overall, we found that the way the labels are used by the training framework is important to preserve the inter-rater variability, while the results of the label fusion methods’ comparison were less univocal.

\subsubsection{When fusing the labels}
Label fusion is a critical step in many image segmentation frameworks as it is often used to condense a collection of labels from multiple raters into a single estimate of the underlying segmentation. Although the GT generated by averaging the raters’ labels is intrinsically a more truthful representation of the inter-rater disagreement than STAPLE (see Figure 2 and Figure 3), training a deep learning model with GT average showed less promising results in this study. The models trained using the averaged GT were underconfident and tended to overestimate the uncertainty, which can be seen on the extended soft edges around the segmented structures (Figure \ref{fig:viz_scgm} and Figure \ref{fig:viz_ms}), the larger underconfidence gaps on reliability diagrams and the associated higher ECE (Figure \ref{fig:cal}), and uncertainty correspondence plots (Figure \ref{fig:unc}). The models trained with labels from individual raters, i.e., random sampling, were less overconfident than when using consensus labels, which is in line with previous studies \citep{jensen2019improving,ji2021learning}. The best calibration results were obtained when using STAPLE as label fusion method, for the MS lesion dataset, and random sampling for SCGM dataset. However, both STAPLE and random sampling had similar reliability diagrams and ECE values (Figure \ref{fig:cal}), suggesting only small differences between “SoftSeg-STAPLE” and “SoftSeg-RandomSampling” candidates in terms of calibration. A similar trend can be observed for the MAE on the uncertainty correspondence plots (Figure \ref{fig:unc}). For both datasets, the Brier score between the GT average and predictions was the best when using random sampling, which is in line with the results obtained by \cite{jungo2018effect}. In terms of segmentation performance, no clear consensus was reached between the two datasets. “SoftSeg-Average” achieved the best performance for MS lesion segmentation, while “SoftSeg-STAPLE” was the best candidate for SCGM. This could be explained by the fact that MS lesion segmentation is more subject to inter-rater disagreement than spinal cord segmentation. MS lesion segmentation models might benefit more from being explicitly exposed to the rater inter-rater variability than the spinal cord segmentation models.  Unlike \cite{jensen2019improving, jungo2018effect, mirikharaji2021d}, no equivocal conclusion can be drawn in terms of the best label fusion method. It would be interesting to extend the study to more datasets to confirm our observations that the more appropriate label fusion method might be dataset-specific.

Ideally, all deep learning models would be trained with GTs derived from multiple raters to account for inter-rater variability. However, the availability of datasets with multiple experts segmenting each image is rare in medical settings, because manual segmentation is time-consuming and expensive. In future works, the impact of having one to N raters in the GT annotations on uncertainty representation should be explored to ensure our results stand with a varying number of raters.

In this work, we focused on three label fusion methods to limit the number of model comparisons. However, more fusion methods exist, such as majority voting, soft STAPLE \citep{kats2019soft}, Bayesian fusion \citep{audelan2022robust}, SIMPLE \citep{langerak2010label}, inter-rater variability sampling scheme \citep{jensen2019improving}, and others. Fusing labels with majority voting is a simpler approach compared to STAPLE and can yield similar performance depending on the number of raters and/or the type of task \citep{mcgurk2013combining}. While the present study did not lead to an unequivocal label fusion method to recommend, some works \citep{audelan2022robust, langerak2010label} suggest methods for improving the STAPLE algorithm, and hence could outperform the fusion methods studied in this work. Future studies could also consider other label fusion methods, such as recently proposed deep-learning approaches to explicitly model the consensus process \citep{nichyporuk2021optimizing,yu2020difficulty}.

\subsubsection{When using the labels through the training pipeline}
The way the labels are processed to train a model has important implications on the preservation of the inter-rater variability in this study. SoftSeg training framework led to a more reliable inter-rater uncertainty and models better calibrated than when using a more conventional training approach. This increased ability to encode the inter-rater variability is probably due to the fact that SoftSeg facilitates the propagation of soft labels throughout the training scheme: (1) no binarization of the input labels, (2) a loss function which does not penalize uncertain predictions, and (3) an activation function which does not enforce binary outputs. SoftSeg has a computational advantage over other uncertainty quantification methods such as the ensembling proposed by \cite{ma2020estimating}. An interesting avenue would be to combine SoftSeg and ensemble, thereby creating an ensemble of SoftSeg models to potentially improve calibration and uncertainty representation. Considered with equivalent expertise in this work, future studies could account for the different expertise across raters, for instance by modulating the training scheme with FiLM layers \citep{lemay2021benefits}, or by the use of expertise-aware inferring module \citep{ji2021learning}.

\subsubsection{When comparing with loss ensembles}
Based on the experiment described in Section \ref{sec:ss-loss-ensembles}, on comparison with an ensembling approach that aimed at better preserving the inter-rater variability and model uncertainties, we observed that SoftSeg-STAPLE performed better in terms of calibrated predictions and uncertainty quantification metrics. This is an important result because not only it suggests that our method is more computationally efficient, i.e., does not require multiple forward passes through each model at test time, unlike loss ensembles \cite{ma2020estimating}). It is also better at preserving inter-rater variability with a single forward pass.

\subsection{A multifaceted evaluation with model training repetitions}
While it is common to select the best model solely based on segmentation accuracy considerations \citep{commowick2018objective,isensee2017brain,prados2017spinal}, we argue that a more exhaustive evaluation is needed, e.g., by including model calibration and uncertainty assessments. For instance, “Conv-STAPLE” is among the best approaches in terms of segmentation accuracy on the MS dataset (see Figure \ref{fig:composite}), but is not properly calibrated as it showed an important overconfident gap (see Figure \ref{fig:cal}). A multifaceted evaluation scheme has the potential to facilitate model acceptance and integration in the clinical routine, which still remains limited \citep{jungo2018effect}. Some avenues are discussed below.

\subsubsection{The ongoing research around uncertainty and calibration estimation}
In the same way that there are numerous segmentation accuracy metrics \citep{yeghiazaryan2015overview}, there are many ways to assess the model uncertainty and calibration. For instance, recent studies suggested the voxel-wise aleatoric \citep{wang2019aleatoric} and epistemic \citep{xia2020uncertainty} uncertainties, or the structure-wise uncertainty \citep{roy2019bayesian}, just to name a few uncertainty evaluation methods. The medical image analysis community has only recently started to report measures of model uncertainty and model calibration, and the best practices on how to estimate them are yet to be determined \citep{abdar2021review, gal2016dropout, guo2017calibration}. We acknowledge the exhaustive comparison performed by \cite{jungo2020analyzing} across different uncertainty estimation methods. Their study showed the limits of voxel-wise uncertainty measures in terms of subject-level calibration and recommended the use of subject-wise uncertainty estimates. We followed their recommendations in the present study. Uncertainty was computed directly from the model prediction, rather than from Monte Carlo iterations or deep ensembles, which does not require more computational power and can be measured during inference. Calibration was qualitatively assessed with reliability diagrams and quantitatively analyzed with the ECE as done by \cite{guo2017calibration}. While multiple studies suggest post-hoc strategies to improve calibration \citep{guo2017calibration, kuleshov2018accurate, zhang2020learning}, we suggest a training strategy that directly generates calibrated outputs without the need of extra computation or hyperparameters. 

\subsubsection{When using the labels through the training pipeline}
With the increased number of evaluation criteria often comes the complexity to select a model as the preferred one. The prioritization of one criterion over the others can be application- or user- specific. Alternatively, in this study, we introduce the use of a composite score to represent the overall segmentation accuracy performance by aggregating multiple scores. This approach assumes equal weights for each evaluation criteria, which can be modified depending on the model user's needs. In lesion segmentation tasks, the overall score proposed by Carass et al. (2017) can also be used in lieu of the composite score used here. Another avenue would be to represent the performance across the different criteria using a radar visualization, e.g., used by \cite{placidi2021multiple}.

\subsubsection{The importance of training repetitions}
Common in studies using machine learning approaches, we observe that experience repetition (e.g., cross-validation, random dataset splittings) is not often performed by studies using deep learning approaches. This is likely due to the long training time required by deep learning model training (often several days). However, our experiments showed that a large variability can be observed across the dataset splittings, especially when data is limited which is often the case in medical settings. For instance, the standard deviation of Dice across the 40 “Conv-STAPLE” models was 12.8\%. In the present study, we performed 40 random dataset splittings for the experiments on the SCGM dataset, and 10 on the MS brain dataset. We encourage future deep learning studies to implement experience repetitions in their evaluation scheme.

\subsection{Limitations}
Since SoftSeg generates softer segmentations, this method is more sensitive to the choice of binarization threshold compared with conventional models where predictions are mostly binary. Hence, thoughtful postprocessing is advised when using SoftSeg. Validation loss progression had different behaviors for conventional and SoftSeg models.  While in the previous \citep{gros2021softseg} and current work, all SoftSeg models converged, the Adaptive Wing loss stagnated for the first 25 epochs on the MS lesion models while the Dice loss gradually decreased during all training (see Appendix A).

\section{Conclusion}
In this study, we evaluated three methods to combine labels from multiple raters using a conventional training framework and SoftSeg, aiming to preserve the inter-rater variability. Our study highlights overconfidence and inter-rater variability underestimation of the conventional framework while SoftSeg models with STAPLE or random sampling were well calibrated and reflected more truthfully the variability due to multiple experts. While fusing annotations using the average encodes the disagreement between experts, predictions were underconfident and the rater uncertainty was overestimated. No consistent observation was made throughout datasets to determine an overall best label fusion method. However, SoftSeg was systematically superior or equal in terms of segmentation performance and had the best calibration and preservation of the inter-rater variability. SoftSeg showed similar results to an ensemble method in terms of uncertainty. SoftSeg has the advantage of requiring the training of a single model and uses a single forward pass while the ensemble requires training of multiple models, one per rater, and requires one forward pass per model. While these observations should be confirmed on other datasets, using SoftSeg could potentially be an effective strategy to capture inter-rater variability in segmentation tasks.


\acks{The authors thank Marie-Hélène Bourget for her methodological insights and Yang Ding, Nick Guenther, Joshua Newton, Ainsleigh Hill, and Alexandru Foias for helping with ivadomed maintenance. Funded by the Canada Research Chair in Quantitative Magnetic Resonance Imaging [950-230815], the Canadian Institute of Health Research [CIHR FDN-143263], the Canada Foundation for Innovation [32454, 34824], the Fonds de Recherche du Québec - Santé [322736], the Natural Sciences and Engineering Research Council of Canada [RGPIN-2019-07244], the Canada First Research Excellence Fund (IVADO and TransMedTech), the Courtois NeuroMod project, the Quebec BioImaging Network [5886, 35450], INSPIRED (Spinal Research, UK; Wings for Life, Austria; Craig H. Neilsen Foundation, USA), Mila - Tech Transfer Funding Program. A.L. has a fellowship from Centre UNIQUE, NSERC and FRQNT. C.G. has a fellowship from IVADO [EX-2018-4], The authors thank the NVIDIA Corporation for the donation of a Titan X GPU.}

%
\ethics{All the data used for this study was de-identified. No IRB was necessary for this work.}

\coi{The authors declare that they have no conflicts of interest including financial interests or personal relationships that could impact the reported results in this paper.}

\bibliography{sample}


\clearpage
\appendix
\section*{Appendix A.}

\begin{figure}[H]
\begin{center}
    \subfloat[\centering SCGM]{\includegraphics[width=0.8\linewidth]{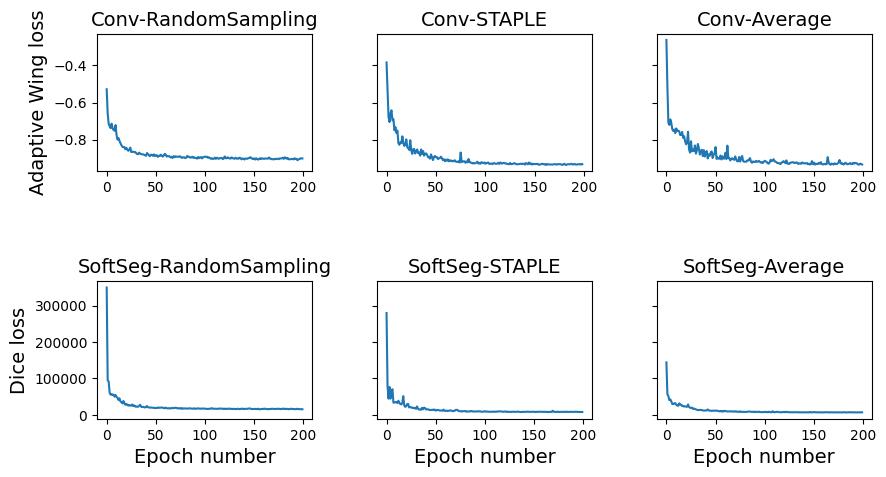} \label{fig:loss_gm}}%
    \qquad
    \subfloat[\centering MS brain]{\includegraphics[width=0.8\linewidth]{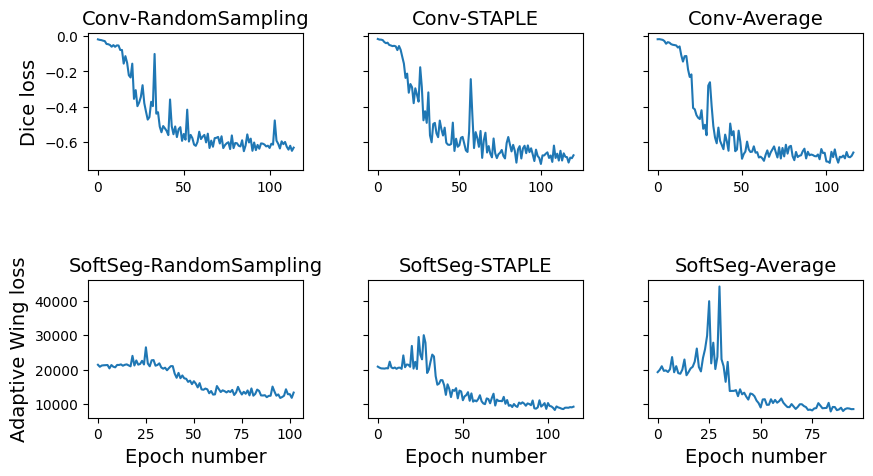} \label{fig:loss_ms}}
\end{center}
\caption{Validation loss progression for a seed from each candidate for the SCGM (a) and MS brain lesions (b) datasets.}
\label{fig:loss}
\end{figure}

\paragraph{Training process}
Table \ref{tab:param} enumerates the training parameters used to train the models from this study. All models were trained for a maximum a 200 epochs with an early stopping of 50 epochs ($\epsilon=0.001$). Figure \ref{fig:loss} illustrates the validation losses progression for on seed from each model. The validation loss reached a plateau for at least the last 50 epochs of the training. The small amount of training data and the high level of difficulty of the MS brain lesion segmentation tasks generated more instability in the validation loss.

\clearpage

\begin{table}[H]
\centering
\caption{Training parameters for each dataset. This table was partially extracted from \cite{gros2021softseg}. Abbreviations: MS: multiple sclerosis; RPI: right-to-left, posterior-to-anterior, inferior-to-superior orientation; SCGM: spinal cord gray matter.}

\makebox[\textwidth][c]{
    
\begin{tabular}{ll|cc}
\cmidrule[\heavyrulewidth]{3-4}
&  & \textbf{SCGM dataset} & \textbf{\thead{Brain MS lesion \\ dataset}} \\
\midrule
\multirow{3}{*}{{\footnotesize{\STAB{\textbf{Preprocessing}}}}} & \textbf{Resample} & \thead{0.25 $\times$ 0.25 $\times$ 2  \\ mm$^3$ (RPI)} & 1 mm isotropic \\ 
& \textbf{Batch format} & \multicolumn{2}{c}{2D axial slices} \\
& \textbf{Crop} & $128 \times 128$ pixels$^2$ & $160 \times 124$ pixels$^2$ \\
\midrule
\multirow{3}{*}{{\footnotesize{\STAB{\textbf{Data augmentation}}}}} & \textbf{Rotation} & \multicolumn{2}{c}{$\pm 20$ degrees}  \\
& \textbf{Translation} & \multicolumn{2}{c}{$\pm 3\%$} \\
& \textbf{Scale} & \multicolumn{2}{c}{$\pm 10\%$} \\
\midrule
\multicolumn{2}{c|}{\textbf{Batch size}} & 8 & 24 \\
\multicolumn{2}{c|}{\textbf{U-Net depth}} & 3 & 4 \\
\multicolumn{2}{c|}{\textbf{Dropout rate}} & \multicolumn{2}{c}{30\%} \\
\midrule
\multirow{2}{*}{{\footnotesize{\STAB{\textbf{Learning Rate}}}}} & \textbf{Initial} & 0.001 & 0.00005  \\
& \textbf{Scheduler} & \multicolumn{2}{c}{Cosine Annealing} \\
\midrule
\multicolumn{2}{c|}{\textbf{Adaptive Wing Loss}} & \multicolumn{2}{c}{$\epsilon=1$; $\alpha=2.1$; $\Omega=0.5$; $\omega=8$} \\
\midrule
\multicolumn{2}{c|}{\textbf{Early stopping}} & \multicolumn{2}{c}{Patience: 50 epochs; $\epsilon: 0.001$} \\
\midrule
\multicolumn{2}{c|}{\textbf{Maximum number of epochs}} & \multicolumn{2}{c}{200} \\

\bottomrule 
\end{tabular}}

\label{tab:param}
\end{table}

\end{document}